\documentclass[prb,preprint,showpacs,preprintnumbers,amsmath,amssymb]{revtex4}
\usepackage{bm}
\usepackage{epsfig}

\begin{document}
\title{Electric-field induced spin excitations in two-dimensional spin-orbit coupled systems}
\author{P. Kleinert}
\affiliation{Paul-Drude-Intitut f\"ur Festk\"orperelektronik,
Hausvogteiplatz 5-7, 10117 Berlin, Germany}
\author{V.V. Bryksin}
\affiliation{A.F. Ioffe Physical Technical Institute,
Politekhnicheskaya 26, 194021 St. Petersburg, Russia}
\date{\today}
\begin{abstract}
Rigorous coupled spin-charge drift-diffusion equations are derived from quantum-kinetic equations for the spin-density matrix that incorporate effects due to ${\bm k}$-linear spin-orbit interaction, an in-plane electric field, and the elastic scattering on nonmagnetic impurities. The explicit analytical solution for the induced magnetization exhibits a pole structure, from which the dispersion relations of spin excitations are identified. Applications of the general approach refer to the excitation of long-lived field-induced spin waves by optically generated spin and charge patterns. This approach transfers methods known in the physics of space-charge waves to the treatment of spin eigenmodes. In addition, the amplification of an oscillating electric field by spin injection is demonstrated.
\end{abstract}


\maketitle

\section{Introduction}
Recent attention has focused on semiconductor spintronics, in which electronic spin polarization is used for information processing. Especially, the generation and manipulation of nonequilibrium spin densities by exclusively electrical means in nonmagnetic semiconductors is particularly attractive. Progress toward the development of spintronic devices depends on theoretical and experimental studies of effects due to the spin-orbit interaction (SOI). This spin-dependent coupling gives rise to an internal effective magnetic field that leads to spin precession and reorientation. For semiconductor quantum wells or heterostructures, the bulk and structural inversion asymmetry give rise to Dresselhaus and Rashba SOI terms, respectively. Unfortunately, the very same SOI also causes spin relaxation. The randomization of electron spins is due to the fact that the SOI depends on the in-plane momentum ${\bm k}$. Consequently, the precession frequencies differ for spins with different wave vectors. This so-called inhomogeneous broadening in conjunction with any scattering causes spin dephasing,\cite{Wu_2001} the details of which depend on the character of dominating scattering processes, the band structure, and the crystal orientation.\cite{JAP_073702} In GaAs/Al$_{x}$Ga$_{1-x}$As quantum wells grown along the [001] axis and with balanced Rashba and Dresselhaus SOI strength, a strong anisotropy in the in-plane spin dephasing time has been measured.\cite{PRB_033305,APL_112111,PRB_073309} The spin relaxation along the [110] direction is efficiently suppressed. Based on this effect, which is robust due to an exact spin rotation symmetry of the spin-orbit Hamiltonian,\cite{PRL_236601} a nonballistic spin-field-effect transistor was proposed.\cite{PRL_146801} From a theoretical point of view, it is predicted that for an idealized model with ${\bm k}$-linear SOI the spin polarization along [110] is conserved for a certain wave vector.\cite{PRL_236601} The experimental confirmation of this prediction \cite{PRL_076604} was possible by exploiting transient spin-grating techniques. This experimental method offers an efficient tool for identifying coupled spin-charge eigenstates in the two-dimensional electron gas (2DEG). Optically induced diffraction patterns are formed in semiconductors, when two pulses with identical energies interfere on the sample and excite electron-hole pairs.\cite{PRL_4793,PRL_136602,PRB_125344,JAP_063714} By varying the relative angle between the two pump beams, the grating period can be tuned for resonant excitation of the eigenmodes. With a third time-delayed pulse that diffracts from the photo-injected spin or charge pattern, the time evolution of the spin polarization is monitored. A free-carrier concentration grating is produced within the sample by two beams with parallel linear polarization. Alternatively, an oscillating spin polarization, which levitates over a homogeneous carrier ensemble, is generated by cross-linearly polarized pump pulses. By detuning the frequencies, a moving (oscillating) charge and/or spin pattern can be produced.

Most interesting both for basic research and the application point of view are weakly damped spin-charge eigenmodes of the semiconductor heterostructure. These excitations drastically change their character when an in-plane electric field acts simultaneously on spin and charge carriers. Similar to the traditional study of space-charge waves in crystals (see, for instance, Ref. \onlinecite{BuchPetrov1}), the field-dependent spin modes can be identified and excited by an experimental set up that provides the appropriate wave vector. Such an approach can profit from methods developed in the field of space-charge waves in crystals.

It is the aim of this paper to systematically derive general spin-charge drift-diffusion equations for a semiconductor quantum well with a general ${\bm k}$-linear SOI. Based on the rigorous analytical solution of these equations, a number of electric-field driven spin resonances are studied.

\section{Solution of drift-diffusion equations}
In this section, we introduce the model, derive and solve general spin-charge coupled drift-diffusion equations for conduction-band electrons in an asymmetric semiconductor quantum well. Coupled spin and charge excitations are treated by an effective-mass Hamiltonian, which includes both SOI and short-range, spin-independent elastic scattering on impurities. We are mainly interested in spin effects exerted by an in-plane electric field ${\bm E}=(E_x,E_y,0)$. The single-particle Hamiltonian
\begin{eqnarray}
H_{0}&=&\sum_{\bm{k},s }a_{\bm{k}s }^{\dag}\left[ \varepsilon_{%
\bm{k}}-\varepsilon _{F}\right] a_{%
\bm{k}s }+\sum_{\bm{k},s ,s ^{\prime }}\left(
{\bm{\Omega}}({
\bm{k}}) \cdot {\bm{\sigma }}_{s s ^{\prime }}\right) a_{\bm{k}%
s }^{\dag}a_{\bm{k}s ^{\prime }}\nonumber\\
&-&e{\bm{E}} \sum_{\bm{k},s}\left. \nabla_{\bm \kappa}
a^{\dag}_{\bm{k}-\frac{\bm{\kappa}}{2}s}a_{\bm{k}+
\frac{\bm{\kappa}}{2}s}\right|_{\bm{\kappa}=\bm{0}}
+u\sum\limits_{\bm{k},
\bm{k}^{\prime}}\sum\limits_{s}a_{\bm{k}s}^{\dag}a_{\bm{k}^{\prime}s},
\label{Hamil}
\end{eqnarray}
is expressed by carrier creation ($a_{\bm{k}s }^{\dag}$) and annihilation ($a_{\bm{k}s }$) operators, the quantum numbers of which are the in-plane wave vector ${\bm k}=(k_x,k_y,0)$ and the spin index $s$. In Eq.~(\ref{Hamil}), $\varepsilon_{\bm k}$, $\varepsilon_F$, and ${\bm \sigma}$ denote the energy of free electrons, the Fermi energy, and the vector of Pauli matrices, respectively. The strength $u$ of elastic scattering can alternatively be characterized by the momentum-relaxation time $\tau$. Contributions of the SOI are absorbed into the definition of the vector ${\bm\Omega}({\bm k})$. We restrict the treatment to linear-in ${\bm k}$ Rashba and Dresselhaus SOI terms, which result from the inversion asymmetry of the quantum-well confining potential and the lack of bulk inversion symmetry. For the combined Rashba-Dresselhaus model, the electric-field induced spin polarization depends on the orientation of the in-plane electric field.\cite{IJMPB_4937,PRB_155323} In addition, spin relaxation and spin transport are not only affected by the orientation of spins, but also by the spin-injection direction.\cite{PRB_205328} In order to be in the position to account for all cases of interest, we consider the general class of ${\bm k}$-linear SOI expressed by $\Omega_{i}({\bm k})=\alpha_{ij}k_{j}$, where $\alpha_{ij}$ are spin-orbit coupling constants.\cite{PRB_205340} The most studied example is a semiconductor quantum well grown along the [001] direction. Under the condition that the Cartesian coordinate axes are chosen along the principal crystallographic directions, we have for the combined Rashba-Dresselhaus model $\alpha_{11}=\beta$, $\alpha_{12}=\alpha$, $\alpha_{21}=-\alpha$, and $\alpha_{22}=-\beta$, with $\alpha$ and $\beta$ being the Rashba and Dresselhaus coupling constants, respectively. A change of the spin-injection direction is achieved by the transformation ${\bm \Omega}^{\prime}=U{\bm \Omega}(U^{-1}{\bm k})$, with $U$ being a rotation matrix.\cite{PRB_205328} A particular configuration is obtained after a rotation around $\pi/4$, which leads to the SOI couplings: $\alpha_{11}=0$, $\alpha_{12}=\alpha-\beta$, $\alpha_{22}=0$, and $\alpha_{21}=-(\alpha+\beta)$.

Spin-related phenomena are completely captured by the spin-density matrix $f_{ss'}({\bm k},{\bm \kappa}|t)$, which is calculated from quantum-kinetic equations.\cite{PRL_226602,PRB_155308,PRB_075306,PRB_165313} Based on the Born approximation for elastic scattering and on low-order corrections of the SOI to the collision integral, Laplace-transformed kinetic equations are obtained for the physical components $f={\rm Tr}\widehat{f}$ and ${\bm f}={\rm Tr}{\bm\sigma}\widehat{f}$ of the spin-density matrix.\cite{PRB_165313} In these equations, spin-dependent contributions to the collision integral are needed to guarantee that the spin system correctly approaches the state of thermodynamic equilibrium. A solution of the coupled kinetic equations is searched for in the long-wavelength and low-frequency regime. Systematic studies are possible under the condition of weak SOI, when a physically relevant evolution period exists, in which the carrier energy is already thermalized, although both the charge and spin densities still remain inhomogeneous. We shall focus on this regime, where the following ansatz for the spin-density matrix is justified:\cite{PRB_075340}
\begin{equation}
\overline{f}({\bm k},{\bm\kappa}|t)=-F({\bm\kappa},t)\frac{dn(\varepsilon_{\bm k})/d\varepsilon_{\bm k}}{d n/d\varepsilon_F},\label{chargef}
\end{equation}
\begin{equation}
\overline{{\bm f}}({\bm k},{\bm\kappa}|t)=-{\bm F}({\bm\kappa},t)\frac{dn(\varepsilon_{\bm k})/d\varepsilon_{\bm k}}{d n/d\varepsilon_F}.\label{spinf}
\end{equation}
The bar over the quantities $f$ and ${\bm f}$ indicates an integration with respect to the polar angle of the vector ${\bm k}$. In Eqs.~(\ref{chargef}) and (\ref{spinf}), $n(\varepsilon_{\bm k})$ denotes the Fermi function and $n=\int d\varepsilon \rho(\varepsilon)n(\varepsilon)$ is the carrier density with $\rho(\varepsilon)$ being the density of states of the 2DEG. By applying the outlined schema, spin-charge coupled drift-diffusion equations are straightforwardly derived for the macroscopic carrier density $F({\bm\kappa},t)$ and magnetization ${\bm M}({\bm\kappa},t)=\mu_B{\bm F}({\bm\kappa},t)$ [with $\mu_B=e\hbar/(2mc)$ being the Bohr magneton]. For the general class of SOI ${\bm\Omega}=\widehat{\alpha}{\bm k}$, we obtain the coupled set of equations
\begin{equation}
\left[\frac{\partial}{\partial t}-i\mu{\bm E}{\bm\kappa}+D{\bm\kappa}^2 \right]F+\frac{i}{\hbar\mu_B}{\bm\Omega}({\bm\kappa}){\bm M}-\frac{2im\tau}{\hbar^3\mu_B}|\widehat{\alpha}|([{\bm\kappa}\times\mu{\bm E}]{\bm M})=0,\label{drift1}
\end{equation}
\begin{equation}
\left[\frac{\partial}{\partial t}-i\mu{\bm E}{\bm\kappa}+D{\bm\kappa}^2+\widehat{\Gamma} \right]{\bm M}-\frac{e}{mc}{\bm M}\times{\bm H}_{eff}-\chi (\widehat{\Gamma}{\bm H}_{eff})\frac{F}{n}-\frac{im\mu^2}{\hbar^2 c}|\widehat{\alpha}|({\bm\kappa}\times{\bm E})F={\bm G},\label{drift2}
\end{equation}
in which the matrix of spin-scattering times $\widehat{\Gamma}$, an effective magnetic field ${\bm H}_{eff}$, and the determinant $|\widehat{\alpha}|$ of the matrix $\widehat{\alpha}$ appear. $\chi=\mu_B^2n^{\prime}$ denotes the Pauli susceptibility. In addition, the spin generation by an external source is treated by the vector ${\bm G}$ on the right-hand side of Eq.~(\ref{drift2}). To keep the representation transparent, let us focus on the combined Rashba-Dresselhaus model $\alpha_{11}=-\alpha_{22}=\beta$, $\alpha_{12}=-\alpha_{21}=\alpha$, for which the coupling parameters are expressed by
\begin{equation}
\alpha=\frac{\hbar^2K}{m}\cos(\psi+\pi/4),\quad \beta=\frac{\hbar^2K}{m}\sin(\psi+\pi/4).
\end{equation}
By choosing $\psi=-\pi/4$ or $\psi=\pi/4$, the pure Rashba or pure Dresselhaus SOI is reproduced, respectively. For the spin scattering matrix, we have
\begin{equation}
\widehat{\Gamma}=\frac{1}{\tau_s}\left(
         \begin{array}{ccc}
           1 & \cos(2\psi) & 0 \\
           \cos(2\psi) & 1 & 0 \\
           0 & 0 & 2 \\
         \end{array}
       \right),
\end{equation}
in which the spin-scattering time $\tau_s$ appears ($1/\tau_s=4DK^2$) with $D$ being the diffusion coefficient of the charge carriers. The electric field enters the drift-diffusion Eqs.~(\ref{drift1}) and (\ref{drift2}) both directly and via an effective magnetic field
\begin{equation}
{\bm H}_{eff}=-\frac{2m^2c}{e\hbar^2}\widehat{\alpha}(\mu {\bm E}+2iD{\bm\kappa}),
\end{equation}
which is related to the SOI. This auxiliary field originates on the one hand from the in-plane electric field ${\bm E}$ and on the other hand from the inhomogeneity of spin and charge densities. For the mobility $\mu$, the Einstein relation holds $\mu=eDn^{\prime}/n$ with $n^{\prime}=dn/d\varepsilon_F$. Variants of Eqs.~(\ref{drift1}) and (\ref{drift2}) have already been published previously.\cite{PRB_052407,PRB_245210,PRL_236601,PRB_041308,PRB_205326}

An exact solution of the spin-charge coupled drift-diffusion Eq.~(\ref{drift2}) for the field-induced magnetization is easily derived for the case ${\bm E}\uparrow\uparrow{\bm\kappa}$. By applying a Laplace transformation with respect to the time variable $t$, we obtain the analytic solution
\begin{equation}
{\bm M}_{\perp}^{\prime}=\frac{\Sigma+\widehat{\Gamma}}{D_T}\left[{\bm Q}_{\perp}-\frac{e}{mc\sigma}Q_z{\bm H}_{eff} \right]-\frac{1}{\sigma D_T}\left(\frac{e}{mc} \right)^2[{\bm H}_{eff}\times({\bm H}_{eff}\times{\bm Q}_{\perp})],\label{Mag1}
\end{equation}
\begin{equation}
M_z=\frac{Q_z}{\sigma}+\frac{e}{mc\sigma}\frac{1}{D_T}{\bm H}_{eff}(\Sigma +\widehat{\Gamma}){\bm Q}_{\perp}-\left(\frac{e}{mc\sigma} \right)^2\frac{1}{D_T}{\bm H}_{eff}(\Sigma+\widehat{\Gamma}){\bf H}_{eff}Q_z,\label{Mag2}
\end{equation}
where ${\bm M}^{\prime}={\bm M}-\chi{\bm H}_{eff}$ and ${\bm M}_{\perp}^{\prime}={\bm e}_z\times{\bm M}^{\prime}$. The inhomogeneity of the transformed Eq.~(\ref{drift2}) for the spin-density matrix is denoted by ${\bm Q}$ and has the form
\begin{equation}
{\bm Q}={\bm M}(t=0)+(i\mu{\bm E}{\bm\kappa}-D{\bm\kappa}^2)\chi{\bm H}_{eff}/s+{\bm G}/s.
\end{equation}
Other quantities that appear in Eqs.~(\ref{Mag1}) and (\ref{Mag2}) are defined by $\Sigma=s-i\mu{\bm E}{\bm\kappa}+D{\bm\kappa}^2$ and $\sigma=\Sigma+2/\tau_s$, with $s$ being the Laplace variable. The general solution in Eqs.~(\ref{Mag1}) and (\ref{Mag2}) provides the basis for the study of numerous spin-related phenomena including effects of oscillating electric fields. Most important is the identification of spin excitations by treating the denominator $D_T$, which is given by
\begin{equation}
D_T=\frac{1}{\sigma}\biggl\{\Sigma\left[\sigma^2+\left(\frac{e}{mc}{\bm H}_{eff} \right)^2 \right]+g^2\left[\sigma+\frac{(\mu{\bm E})^2}{D} \right] \biggl\},\label{det}
\end{equation}
with $g=4Dm^2|\widehat{\alpha}|/\hbar^4$. The cubic equation $D_T=0$ with respect to the Laplace variable $s\rightarrow -i\omega$ yields three spin-related eigenmodes that have already been studied for zero electric field ${\bm E}={\bm 0}$ in Ref.~\onlinecite{PRB_125307}. Field-dependent eigenstates calculated from the zeros of Eq.~(\ref{det}) are characterized both by the direction of the electric field and  by the spin injection/diffusion direction. Most solutions of this equation describe damped resonances in the charge transport and spin polarization. However, as already mentioned, there also exist undamped excitations, which have received particular interest in recent studies. Here, we focus on these long-lived spin states and study their dependence on an in-plane electric field.

\section{Study of field-dependent eigenmodes}
In the second part of the paper, we present selected applications of the general approach presented in the previous section. We focus on spin effects on charge transport exerted by spin injection and on long-lived field-induced spin excitations probed by interference gratings.

\subsection{Amplification of an electric field by spin injection}
As a first application of our general approach, the charge-current density is studied on the basis of its definition
\begin{equation}
{\bm j}(t)=-ie\left. \nabla_{\bm \kappa}F({\bm\kappa},t)\right|_{\bm\kappa=0}.
\end{equation}
Taking into account Eq.~(\ref{drift1}) and the solution in Eqs.~(\ref{Mag1}) and (\ref{Mag2}), the components of the conductivity tensor are straightforwardly calculated. Here, we are interested in combining the ac electric field with an external permanent spin-injection source that provides a generation rate $G_z(s)$ for the out-of-plane spin polarization.

\begin{figure}
\centerline{\epsfig{file=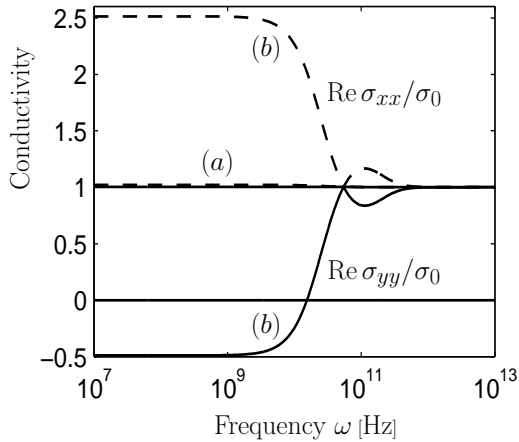,width=7.0cm}}
\caption{Real part of the diagonal conductivity components $\sigma_{xx}$ (dashed lines) and $\sigma_{yy}$ (solid lines) as a function of frequency $\omega$ for $\beta/\alpha=0.5$, $\alpha=10^{-9}$~eVcm, $\tau=10^{-13}$~s, and $D=24$~cm$^2$/s. The sets of dashed and solid lines (a) and (b) are calculated with $G_{z0}/n=0$ and $0.03$, respectively.}
\label{Figure1}
\end{figure}

Focusing on the linear response regime with respect to the in-plane electric field, an analytical expression for the conductivity tensor $\widehat{\sigma}$ is obtained. In accordance with the applicability of the drift-diffusion approach, the derivation is restricted  to the case, when the inequality $\hbar n^{\prime}/(\tau n)\ll 1$ is satisfied. For the frequency-dependent ($s\rightarrow -i\omega$) longitudinal and Hall conductivities, we obtain
\begin{equation}
\sigma_{\stackrel{xx}{yy}}(s)/\sigma_0=1\pm \frac{G_z\tau_s}{n}\frac{\hbar n^{\prime}}{4\tau n} \frac{\sin(4\psi)}{(s\tau_s+1)(s\tau_s+2\cos^2\psi)(s\tau_s+2\sin^2\psi)}
\end{equation}
\begin{equation}
\sigma_{\stackrel{xy}{yx}}(s)/\sigma_0=\pm \frac{G_z\tau_s}{n}\frac{\hbar n^{\prime}}{2\tau n}
\frac{\sin(2\psi)}{s\tau_s+2}\left[\frac{\tau}{\tau_s}-\frac{s\tau_s+1}{(s\tau_s+2\cos^2\psi)(s\tau_s+2\sin^2\psi)} \right],
\end{equation}
with $\sigma_0=e\mu n$. Other contributions to the charge transport, which are solely due to SOI, are much weaker than the retained terms originating from spin injection. This conclusion is illustrated by the curves (a) in Fig.~\ref{Figure1}, which have been numerically calculated for the case $G_z=0$. Weak SOI leads only to a slight deviation of the longitudinal conductivities $\sigma_{xx}$ and $\sigma_{yy}$ from $\sigma_0$. The spin effect completely disappears for the special Rashba-Dresselhaus model with $\alpha=\beta$ ($\psi=0$). The situation drastically changes, when there is an appreciable permanent spin injection, which leads to additional contributions to the steady-state charge transport owing to the spin-galvanic effect. An example is shown by the curves (b) in Fig.~\ref{Figure1}. The striking observation is that Re~$\sigma_{yy}$ becomes negative for frequencies below about $10^{10}$~Hz. This remarkable behavior is confirmed by the expression for the static conductivity
\begin{equation}
\sigma_{\stackrel{xx}{yy}}(s)/\sigma_0=1\pm \frac{G_z\tau_s}{n}\frac{\hbar n^{\prime}}{4\tau n}\cot(2\psi),
\end{equation}
from which it is obvious that $\sigma_{yy}$ changes its sign for sufficiently strong spin injection. Therefore, we meet the particular situation that a paramagnetic medium, which usually absorbs energy from an ac electric field to produce a
spin accumulation, is driven to another regime, where the ac field, which propagates in a given direction, is amplified by spin injection. This stimulated emission is similar to the microwave energy gain, which was recently predicted to occur in a paramagnetic medium with sufficiently large injection spin currents.\cite{PRL_116601} Based on their findings derived for rotating magnetic fields, the authors proposed a new concept for a spin-injection maser. Our result is in line with their conclusion.

\subsection{Electric-field mediated spin excitations}
\begin{figure}
\centerline{\epsfig{file=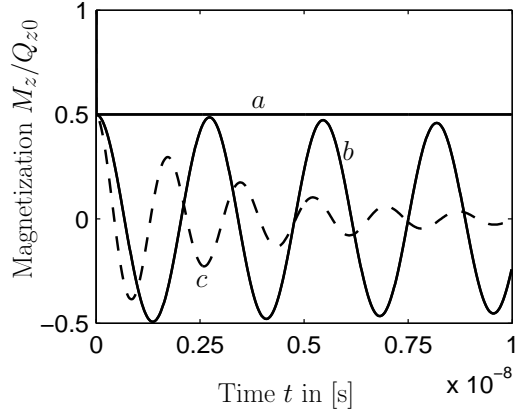,width=7.0cm}}
\caption{Time dependence of the electric-field induced out-of-plane magnetization for $E_{+}=500$~V/cm, $D=24$~cm$^2$/s, and $\tau=10^{-13}$~s. The lines a, b, and c were calculated for ($\kappa_{+}/\sqrt{2}K=2$, $\alpha=\beta=10^{-9}$eVcm), ($\kappa_{+}/\sqrt{2}K=2.01$, $\alpha=\beta=10^{-9}$eVcm), and ($\kappa_{+}/\sqrt{2}K=2.02$, $\alpha=1.05$, $\beta=0.95\,10^{-9}$eVcm), respectively. In addition, we set $r_{+}=0$.}
\label{Figure2}
\end{figure}

As a second application of our general approach, we treat coupled spin-charge eigenstates that exist in a biased sample. Effects of this kind depend not only on the directions of the electric field and the spin injection, but also on the orientation of the crystallographic axes. Here, we study the influence of an electric field on an optically generated standing spin lattice that is periodic along the $\kappa_{+}=(\kappa_x+\kappa_y)/\sqrt{2}$ direction. For simplicity, it is assumed that the spin generation provides a regular lattice for the out-of-plane spin polarization
\begin{equation}
Q_z(\kappa)=\frac{Q_{z0}}{2}[\delta(\kappa_{+}-\kappa_0)+\delta(\kappa_{+}+\kappa_0)].
\end{equation}
Inserting this source term into Eq.~(\ref{Mag2}), we obtain for the related field-dependent magnetization the solution
\begin{equation}
M_z(\kappa,s)=\frac{\Sigma\sigma +g^2}{\Sigma\left[\sigma^2+\left(\frac{e}{mc}{\bm H}_{eff} \right)^2 \right]+g^2\left[\sigma +(\mu{\bm E})^2/D \right]}Q_z(\kappa),\label{GlMz}
\end{equation}
the character of which is mainly determined by the poles calculated from the zeros of the denominator. Pronounced oscillations arise for the special Rashba-Dresselhaus model with $\alpha=\beta$. In this case ($g=0$), Eq.~(\ref{GlMz}) is easily transformed to spatial and time variables with the result
\begin{eqnarray}
M_z(r_{+},t)=\frac{Q_{z0}}{2}&\biggl\{&e^{-D(\kappa_0+2\sqrt{2}K)^2t}\cos(\kappa_0r_{+}+\mu E_{+}(\kappa_0+2\sqrt{2}K)t)\nonumber\\
&+&e^{-D(\kappa_0-2\sqrt{2}K)^2t}\cos(\kappa_0r_{+}+\mu E_{+}(\kappa_0-2\sqrt{2}K)t)\biggl\},
\end{eqnarray}
where $K=\sqrt{2}m\alpha/\hbar^2$ and $r_{+}=(r_x+r_y)/\sqrt{2}$. In general, this solution describes damped oscillations of the magnetization. However, due to a spin-rotation symmetry, there appears an undamped soft mode, when the wave-vector component $\kappa_0$ of the imprinted spin lattice matches the quantity $2\sqrt{2}K$, which is a measure of the SOI. This eigenmode leads to long-lived oscillations of the magnetization. Numerical results, calculated from Eq.~(\ref{GlMz}), are shown in Fig.~\ref{Figure2}.
Under the ideal condition $\alpha=\beta$ and $\kappa_0=2\sqrt{2}K$, the induced magnetization rapidly reaches the value $M_z=Q_{z0}/2$ (curve a) and remains constant afterwards. However, any slight detuning of this special set of parameters sparks damped oscillations that can last for many nano-seconds. Examples are shown by the curves b and c in Fig.~\ref{Figure2}. The importance of such spin-coherent waves, especially their potential for future spintronic applications, has recently been emphasized by Pershin.\cite{PRB_155317} The long-lived spin waves that have been examined in this section are solely generated by an in-plane electric field. We see in them building blocks for future spintronic device applications that rely exclusively on electronic means for generating and manipulating spin.

\section{Excitation of spin waves}
Another wide research area that is covered by the spin-charge coupled drift-diffusion Eqs.~(\ref{drift1}) and (\ref{drift2}) [or their solution in Eqs.~(\ref{Mag1}) and (\ref{Mag2})] refers to the response of the spin subsystem to space-charge waves in semiconductor nanostructures. To provide an example for this kind of studies, we focus in this section on spin waves that are excited by an optically induced moving charge-density grating. Two laser beams with a slight frequency shift between them produce a moving interference pattern on the surface of the semiconductor sample that leads to a periodic generation rate for electron and holes of the form
\begin{equation}
g(x,t)=g_0+g_m\cos(K_g x-\Omega t),\label{gen}
\end{equation}
with a homogeneous part $g_0$ and a modulation $g_m$. $K_g$ and $\Omega$ denote the wave vector and frequency of the grating. The generation rate $g(x,t)$ causes electron [$F(x,t)$] and hole [$P(x,t)$] density fluctuations that have the same spatial and temporal periodicity as the source $g(x,t)$. The dynamics of photogenerated electrons and holes is described by continuity equations, which encompass both carrier generation [$g(x,t)$] and recombination [$r(x,t)$] as well as drift and diffusion (see, for example, Ref. \onlinecite{APL_073711}). If the retroaction of spin on the carrier ensemble is neglected, we obtain the set of equations
\begin{equation}
\frac{\partial F}{\partial t}=\frac{1}{e}\frac{\partial J_n}{\partial x}+g(x,t)-r(x,t),
\end{equation}
\begin{equation}
\frac{\partial P}{\partial t}=-\frac{1}{e}\frac{\partial J_p}{\partial x}+g(x,t)-r(x,t)
\end{equation}
where the current densities for electrons [$J_n(x,t)$] and holes [$J_p(x,t)$] are calculated from drift and diffusion contributions
\begin{equation}
J_n=e\mu_nE_{x}F+eD_n\frac{\partial F}{\partial x},
\end{equation}
\begin{equation}
J_p=e\mu_pE_{x}P-eD_p\frac{\partial P}{\partial x}.
\end{equation}
In these equations, $\mu_n$ and $\mu_p$ ($D_n$ and $D_p$) denote the mobilities (diffusion coefficients) for electrons and holes, respectively. A constant electric field $E_0$ applied along the $x$ direction is complemented by a space-charge field $\delta E(x,t)$, which is calculated from unbalanced electron and hole densities via the Poisson equation
\begin{equation}
\frac{\partial E_{x}}{\partial x}=\frac{4\pi e}{\varepsilon}(P-F),\label{Poiss}
\end{equation}
with $\varepsilon$ being the dielectric constant. The optical grating leads to a weak modulation of the carrier densities around their mean value ($F=F^{0}+\delta F$, $P=P^{0}+\delta P$). Due to the spin-charge coupling manifest in Eqs.~(\ref{drift1}) and (\ref{drift2}), charge-density waves are transferred to the spin degrees of freedom and vice versa. As the hole spin relaxation is rapid, the time evolution of the generated spin pattern can be interpreted in terms of the motion of electrons alone. Consequently, the hole density is not considered in the equations for the magnetization.

In the absence of the optical grating, there is no out-of-plane spin polarization ($F_z^{0}=0$). For the in-plane components, a short calculation leads to the result
\begin{equation}
F_x^{0}=-\frac{1}{2}\hbar K_{11}\mu_nE_0n^{\prime},
\end{equation}
\begin{equation}
F_y^{0}=-\frac{1}{2}\hbar K_{21}\mu_nE_0n^{\prime},
\end{equation}
which expresses the well-known effect of the electric-field mediated spin accumulation.~\cite{Edelstein,Aronov} In these equations, the spin-orbit coupling constants are denoted by $K_{ij}=2m\alpha_{ij}/\hbar^2$. Besides this homogeneous spin polarization, there is a field-induced contribution, which is due to the optical grating. For the respective spin modulation, the harmonic dependence of the carrier generation in Eq.~(\ref{gen}) via $z=K_gx-\Omega t$ remains intact. In view of the periodic boundary condition that we naturally accept for the optically induced grating, it is expedient to perform a discrete Fourier transformation with respect to the $z$ variable according to the prescription $F(z)=\sum_{p}\exp(ipz)F(p)$. The resulting equations for the Fourier coefficients of the magnetization are easily solved by perturbation theory with respect to the optically-induced electric field $Y=\delta E/E_0$ and spin $\delta {\bm F}$ contributions. For the field-dependent homogeneous spin components ($p=0$), we obtain the solution
\begin{equation}
\delta F_x(0)=\frac{2K_{12}n}{g\tau_s eE_0 n^{\prime}}\delta F_z(0),\label{Fx0}
\end{equation}
\begin{equation}
\delta F_y(0)=\frac{2K_{22}n}{g\tau_s eE_0 n^{\prime}}\delta F_z(0),\label{Fy0}
\end{equation}
\begin{equation}
\delta F_z(0)=-\frac{(\mu_n E_0)^2/D_n}{2/\tau_s+(\mu_n E_0)^2/D_n}\biggl\{S(1)Y(-1)+S(-1)Y(1) \biggl\},\label{Fz0}
\end{equation}
in which the $p=1$ spin fluctuation occurs via the quantity
\begin{equation}
S(1)=\delta F_z(1)+\frac{\tau_s}{2\tau_E}\frac{K_{11}}{K_g}\delta F_y(1)-\frac{\tau_s}{2\tau_E}\frac{K_{21}}{K_g}\delta F_x(1).\label{S1}
\end{equation}
The scattering time $\tau_{E}$, which is provided by the constant electric field, is given by $1/\tau_{E}=\mu_n E_0K_g$. The spin response described by Eqs.~(\ref{Fx0}) to (\ref{S1}) is a consequence of electric-field fluctuations that accompany the optically induced charge modulation. As we neglect the retroaction of the induced spin fluctuation on the charge balance, the determination of $Y(p=1)$ rests exclusively on Eqs.~(\ref{gen}) to (\ref{Poiss}). The calculation has been performed in our previous work.\cite{APL_073711} To keep our presentation self-contained, we present previously derived results that are needed for the calculation of the spin polarization. The relative electric-field modulation, which has the form
\begin{equation}
Y(1)=-\frac{g_m}{2g_0}\frac{1+i\Lambda_{-}}{\tau\tau_M(\Omega-\Omega_1)(\Omega-\Omega_2)},
\end{equation}
exhibits characteristic resonances at eigenmodes of space-charge waves given by
\begin{equation}
\Omega_{1,2}=-\frac{1}{2}(\mu_{-}E_0K_g+i\Gamma)\pm\sqrt{\left(\frac{1}{2}(\mu_{-}E_0K_g+i\Gamma) \right)^2+(1+\alpha_1)/\tau\tau_M}.\label{Om12}
\end{equation}
The damping of this mode
\begin{equation}
\Gamma=D_{+}K_g^2+\frac{1}{\tau_M}+\frac{1}{\tau}
\end{equation}
includes the Maxwellian relaxation time $\tau_M=\varepsilon/4\pi\sigma_d$ with $\sigma_d=eg_0\tau\mu_{+}$ and $\mu_{\pm}=\mu_n\pm\mu_p$. The parameter $\alpha_1$ in Eq.~(\ref{Om12}) depends on the electric field and is calculated from
\begin{equation}
\alpha_1=d_{+}(\Lambda_{+}-\mu\Lambda_{-})+\kappa d_{+}(1-\mu^2+\Lambda_{+}^2-\Lambda_{-}^2)+\kappa\Lambda_{+},
\end{equation}
where $d_{\pm}=\mu_{\pm}E_0K_g\tau/2$, $\Lambda_{\pm}=D_{\pm}K_g/\mu_{+}E_0$, $\kappa=(\varepsilon/4\pi e)E_0K_g/(2g_0\tau)$, and $\mu=\mu_{-}/\mu_{+}$. The resonant amplification of dc and ac current components due to space-charge waves provides information useful for the determination of the lifetime and the mobilities of photo-generated electrons and holes in semiconductors.\cite{APL_073711}

To continue the analysis of the spin response, the set of linear equations for the $p=1$ Fourier coefficients of the spin vector must be solved. The analytic solution has the form
\begin{eqnarray}
S(1)=&&\frac{\hbar\tau_s n^{\prime}}{4\tau_E^2K_g^2}\frac{g\tau_E}{\widetilde{D}_T}\left[1+\frac{1+2i\Lambda}{1+\widetilde{\Sigma}\tau_s/2\tau_E} \right](K_{11}K_{12}+K_{21}K_{22})\nonumber\\
&&\times\biggl\{\widetilde{\Sigma}(1+2i\Lambda)\frac{\delta F(1)}{n}+(\widetilde{\Sigma}-i)Y(1) \biggl\},
\label{SS1}
\end{eqnarray}
where $\widetilde{\Sigma}=\Lambda-i(1+\Omega\tau_E)$ and $\Lambda=D_nK_g/\mu_nE_0$. Again, the denominator $\widetilde{D}_T$ in Eq.~(\ref{SS1}) is used for the identification of electric-field-induced eigenmodes of the spin system. For the specific set-up treated in this section, the general expression given in Eq.~(\ref{det}) reduces to the dimensionless form
\begin{eqnarray}
\widetilde{D}_T=\frac{1}{\widetilde{\Sigma}+2\tau_E/\tau_s}
&&\biggl\{\widetilde{\Sigma}\left[(\widetilde{\Sigma}+2\tau_E/\tau_s)^2+\frac{K_{11}^2+K_{21}^2}{K_g^2}(1+2i\Lambda)^2 \right]
\nonumber\\
&&+(g\tau_E)^2\left[\widetilde{\Sigma}+\frac{2\tau_E}{\tau_s}+\frac{(1+2i\Lambda)^2}{\Lambda} \right]\biggl\}.
\label{detneu}
\end{eqnarray}
The closed solution in Eqs.~(\ref{Fx0}) to (\ref{detneu}) for the homogeneous spin polarization, which is due to a moving optical grating, has a resonant character, when eigenmodes of the spin subsystem are excited. The dispersion relations of these modes are obtained from the cubic equation $\widetilde{D}_T=0$ with respect to $\Omega$. Depending on the relative strength of the imaginary part in the equation $\Omega=\Omega(\kappa)$, a more or less pronounced resonance occurs in the induced spin polarization. Due to the spin-rotation symmetry of the model, the special Rashba-Dresselhaus system with $\alpha=\beta$ provides an attractive example. By rotating the spin-injection direction, the situation becomes even more interesting. A rotation around $\pi/4$ leads to the set of SOI parameters $\alpha_{11}=\alpha_{12}=\alpha_{22}=0$, and $\alpha_{21}=-2\alpha$. In this special case ($g=0$), we obtain for the $p=1$ Fourier component of the in-plane spin polarization the result
\begin{equation}
\delta F_y(1)=\frac{i\hbar n^{\prime}}{2\tau_E}\frac{K_{21}}{K_g}\frac{\Lambda\left[1+(K_{21}/K_g)^2\right]-i(1+\Omega\tau_E)}
{\tau_E^2(\Omega+\Omega_{s1})(\Omega+\Omega_{s2})}Y(1),
\end{equation}
in which two field-induced spin eigenmodes appear, the dispersion relations of which are expressed by
\begin{equation}
\Omega_{s1,2}=\mu_n E_0(K_g\mp K_{21})+iD_n(K_g\mp K_{21})^2.
\end{equation}
\begin{figure}
\centerline{\epsfig{file=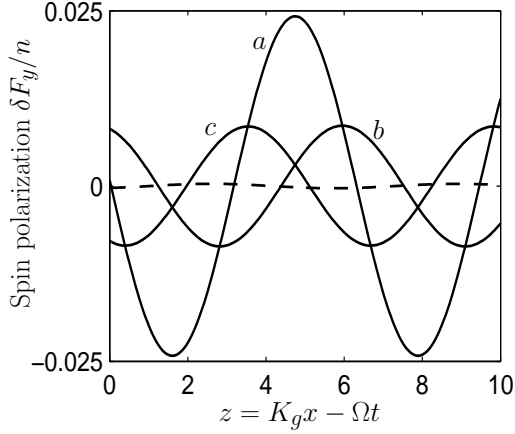,width=7.0cm}}
\caption{Induced in-plane spin polarization as a function of $z=K_gx-\Omega t$ for $\alpha=\beta=0.2\times 10^{-9}$~eVcm, $g_0=g_m=10^{19}$~cm$^{-3}$s$^{-1}$, $\mu_n=0.5$~cm$^2$/Vs, $\mu_p=0.2$~cm$^2$/Vs, $T=77$~K, and a recombination time $\tau_r=10^{-6}$~s. The curves $a$, $b$, and $c$ are calculated with $E_{0max}=2.67$~kV/cm, $E_{0max}+100$~V/cm, and $E_{0max}-100$~V/cm, respectively. For the dashed line, we used $E_0=1$~kV/cm.}
\label{Figure3}
\end{figure}

Again the damping of one mode completely disappears under the condition $K_g=-K_{21}=4m\alpha/\hbar^2$. This peculiarity gives rise to a pronounced resonance in the field dependence of the spin dynamics at $\mu_n E_{0max}=\Omega/(K_g+K_{21})$. Figure~\ref{Figure3} illustrates this effect. Calculated is the in-plane spin modulation $\delta F_y(K_gx-\Omega t)=2{\rm Re}\,\delta F_y(1)\exp(iz)$. The smooth dashed line displays the response of the spin polarization to the optically generated moving charge-density pattern for $E_0=1$kV/cm. The unpretentious signal is considerably enhanced under the resonance condition, when $E_0=E_{0max}=2.67$kV/cm (curve a in Fig.~\ref{Figure3}). By changing the field strength a little bit [$E_0=E_{0max}+100$V/cm (curve b), $E_0=E_{0max}-100$V/cm (curve c)], the phase, amplitude, and frequency of the spin wave drastically change. This resonant influence of an electric field on the excited spin waves is a pronounced effect that is expected to show up in experiments. By applying a magnetic field, the resonant in-plane spin polarization is rotated to generate an out-of plane magnetization.

\section{Summary}
The generation and manipulation of a spin polarization in nonmagnetic semiconductors by an electric field is a subject that has recently received considerable attention. All information needed for the description of these field-induced spin effects are given by the spin-density matrix, the equation of motion of which is governed by the model Hamiltonian. The SOI is the main ingredient in this approach. In spite of the fact that kinetic equations for the four-component spin-density matrix are straightforwardly derived, at least two subtle features have to be taken into account: (i) for the consistent treatment of scattering, its dependence on SOI must be considered, and (ii) to reproduce the well-known field-induced spin accumulation, third-order spin corrections have to be retained in the kinetic equations. In order to study macroscopic spin effects, it is expedient to suppress still existing superfluous information in the spin-density matrix by deriving spin-charge coupled drift-diffusion equations. Under the condition of weak SOI, we followed this program and derived Eqs.~(\ref{drift1}) and (\ref{drift2}) in an exact manner. These equations, which are valid for the general class of linear SOI, apply to various electric-field-induced spin effects that depend not only on the orientation of the crystallographic axes, but also on the spin-injection direction and the alignment of the in-plane electric field. An exact solution of the basic equations for the magnetization allows the identification of field-dependent spin excitations. Among these spin eigenmodes there are long-lived spin waves that can be excited by a spin and/or charge grating providing the necessary wave vector. The applicability of our general approach was illustrated by a few examples. The treatment of the spin-mediated conductivity of charge carriers reveals the possibility that a component of an ac electric field is amplified by spin injection. A similar effect led to the recent proposal for a spin-injection maser device.\cite{PRL_116601} In a second application, it was demonstrated how a regular lattice of an out-of-plane spin polarization excites long-lived field-dependent spin waves. The calculation refers to a [001] semiconductor quantum well with balanced Rashba and Dresselhaus SOI, for which a persistent spin helix has been identified.\cite{PRL_236601} Our final example establishes a relationship to the rich physics of space-charge waves. By considering a typical set-up for the optical generation of a moving charge pattern, the associated dynamics of the related spin degrees of freedom was treated. It was shown that the charge modulation can be used to excite intrinsic field-dependent spin waves. This example demonstrates that the powerful methods developed in the field of space-charge waves can be transferred to the study of spin excitations.

\section*{Acknowledgements}
Partial financial support by the Deutsche Forschungsgemeinschaft
is gratefully acknowledged.



\end{document}